\documentclass[conference]{IEEEtran}
\IEEEoverridecommandlockouts
\usepackage{cite}
\usepackage{amsmath,amssymb,amsfonts}
\usepackage{algorithmic}
\usepackage{algorithm}
\usepackage{graphicx}
\usepackage{textcomp}
\usepackage{xcolor}
\usepackage{multirow}
\usepackage{url} 
\def\BibTeX{{\rm B\kern-.05em{\sc i\kern-.025em b}\kern-.08em
    T\kern-.1667em\lower.7ex\hbox{E}\kern-.125emX}}

\author{
    \IEEEauthorblockN{Zhongli Fang\textsuperscript{1,2},
    Yiran Chen\textsuperscript{1,2},
    Lingyun Zhang\textsuperscript{1,2},
    Yu Liu\textsuperscript{1,2},
    Ping Chen\textsuperscript{2}*,
    Xiaoyan Sun\textsuperscript{3}* and Jun Dai\textsuperscript{3}*
    \thanks{
    *Corresponding authors: Ping Chen (pchen@fudan.edu.cn),   Xiaoyan Sun (xsun7@wpi.edu) and Jun Dai (jdai@wpi.edu)
    }}
    \IEEEauthorblockA{
    \textsuperscript{1}School of Computer Science and Technology, Fudan University, China \\
  \textsuperscript{2}Institute of Big Data, Fudan University,  China\\
  \textsuperscript{3} Worcester Polytechnic Institute
    }
}

\begin{document}

\title{A Trustworthy Watermarking Framework for LLM-Generated Food Safety Content}


\maketitle

\begin{abstract}

Large language models are transforming many industries with their text generation abilities. However, their outputs can be easily tampered with, creating serious risks in critical areas such as food safety reporting. To protect the integrity and traceability of AI-generated content, this paper introduces ToSS (Token Oriented Repartitioning and Strategic Selection), a reliable authentication method using adaptive dual watermarking. The key innovation of ToSS is its dual watermark encoding approach that divides vocabulary tokens into black and white sublists, enabling precise bit-level embedding of traceability information. Additionally, an entropy adaptive mechanism dynamically selects text regions with high prediction uncertainty for watermark insertion, maintaining text fluency and factual accuracy while ensuring reliable traceability. Experiments on multiple datasets, including food domain texts, demonstrate that ToSS achieves leading performance in both watermark capacity and decoding accuracy.



\end{abstract}


\section{Introduction}
\label{sec:intro}





With the rapid development of artificial intelligence, various industries are accelerating their transition towards intelligentization \cite{rombach2022high}. The rise of generative models, particularly large language models (LLMs) \cite{naveed2025comprehensive}, has significantly advanced technologies in speech, image, and text generation, profoundly changing people's lives and work patterns. However, the swift pace of technological advancement has also introduced significant regulatory and security challenges. In recent years, the misuse of generative models to create fake news, celebrity images, and tamper with critical documents has become increasingly prominent. Especially in the food sector, LLMs could be maliciously used to forge food safety inspection reports, alter nutritional composition tables, generate false traceability information, or even fabricate misleading nutritional advice or fake health certifications, posing serious threats to public health and market order. Therefore, there is an urgent need for effective content authentication mechanisms to prevent such misuse. Watermarking technology, as a viable solution, can embed traceable identifiers into generated content. By detecting the presence of a watermark (zero-bit watermarking) or extracting embedded specific information (multi-bit watermarking), it enables authentication of content origin and user identity tracing, thereby ensuring content authenticity and trustworthiness.

Current works, such as the zero-bit watermarking scheme by Kirchenbauer et al. \cite{kirchenbauer2023watermark}, have made significant progress in proving the existence of generated text, while Qu et al. \cite{qu2025provably} have explored the embedding of multi-bit information. However, these existing approaches generally struggle to achieve stable embedding and reliable extraction of long-bit information within limited text lengths while preserving text quality. Particularly in practical application scenarios like food reports and nutritional guidance, there is often a need to embed complete traceability information, certification numbers, or timestamps as long-bit data within a single paragraph or short statement \cite{ji2023survey}. The performance of existing technologies in such short texts remains insufficient. Consequently, there is still a lack of watermarking techniques capable of effectively storing long-bit information within a single paragraph, which limits their practical application in fine-grained authentication scenarios.

\begin{figure}[t] 
\centering
  \includegraphics[width=0.8\linewidth]{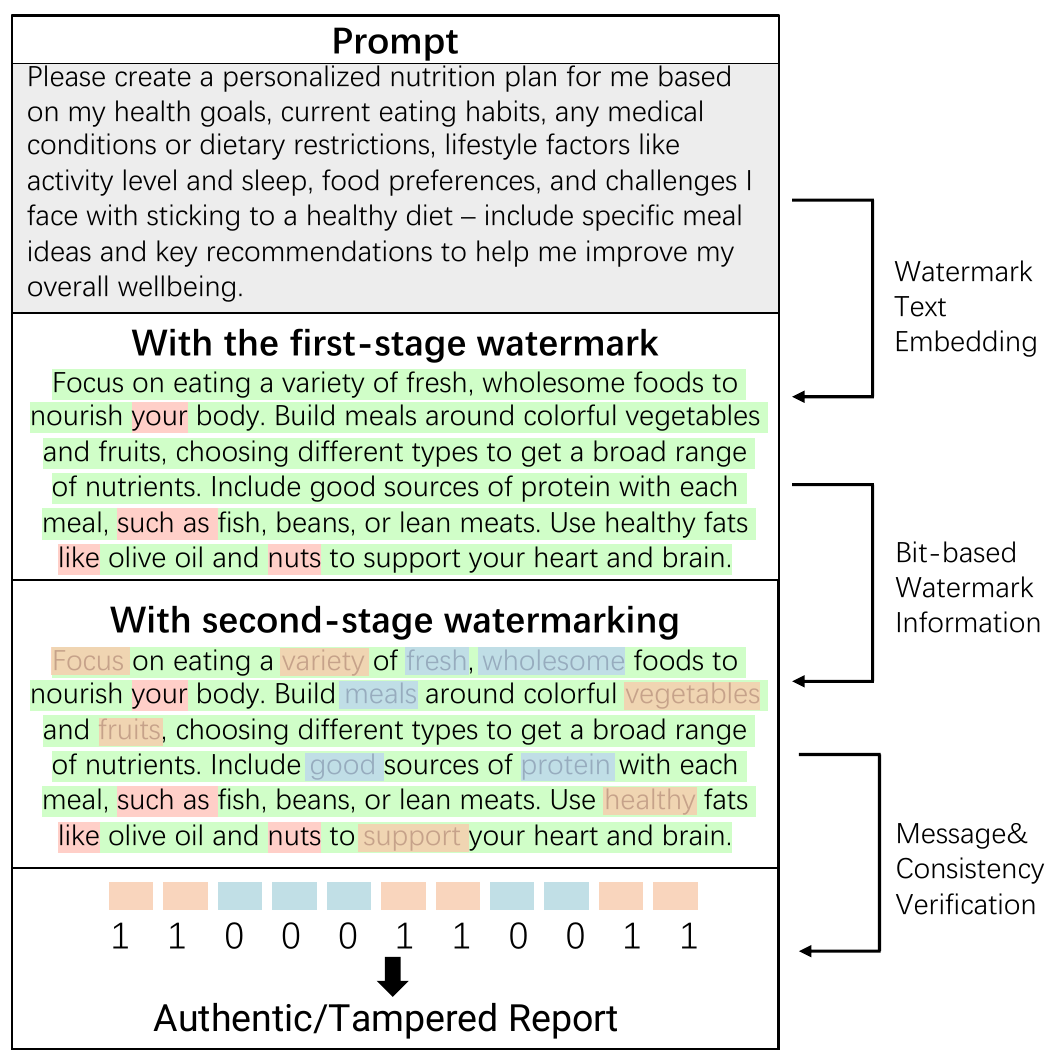} 
  \caption{Illustration of watermark extraction in the ToSS framework.
The first-stage watermark verifies existence using red and green token lists, while the second-stage watermark encodes multi-bit traceability information through black (orange) and white (blue) token lists.
The extracted bit sequence is compared with the stored reference to determine whether the text is authentic or tampered.}   
  \label{fig1}
\end{figure}

\begin{figure*}[t] 
\centering
  \includegraphics[width=0.65\linewidth]{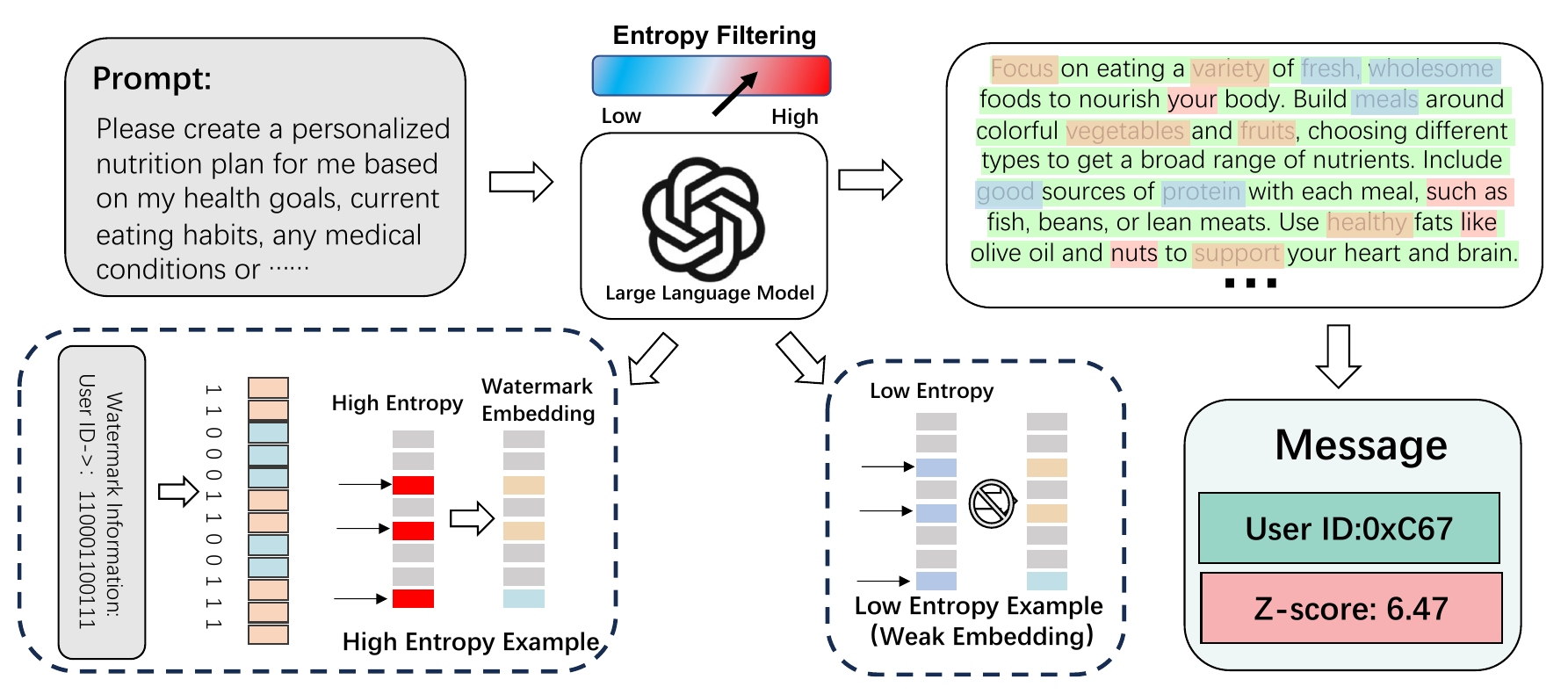} 
  \caption{Overview of the ToSS framework. Given an input prompt, the language model generates text while the entropy-based filtering module identifies high-entropy and low-entropy regions. Multi-bit watermark information is embedded only in high-entropy regions via dynamic black–white list selection, while low-entropy regions are preserved to maintain textual fluency and factual accuracy.}   
  \label{fig2}
\end{figure*}

In contrast, this paper proposes an adaptive dual-watermarking framework named Token-Oriented Repartitioning and Strategic Selection(ToSS) for authenticating content generated by large models, as illustrated in Fig. \ref{fig1}. In the dual‑watermark design, the first-stage watermark partitions the vocabulary into a green list and a red list, and biases generation toward green‑list tokens to enable zero-bit watermark existence verification.
The second-stage watermark further divides the green list into a white list (blue tokens) and a black list (orange tokens), embedding multi-bit traceability information through forced sampling according to the target bit sequence. In addition, ToSS incorporates an entropy‑adaptive mechanism: watermark embedding is strengthened in high‑entropy regions, while only a light green‑list bias is applied in low‑entropy regions to preserve textual fluency.

During verification, the system first performs zero‑bit detection and then reconstructs the multi-bit sequence using the black/white list assignments, comparing it against the stored reference to determine whether the text has been tampered with. Experimental results show that ToSS can robustly embed and extract watermarks of up to 64 bits in short‑text scenarios such as food safety reports and nutritional guidance, and significantly outperforms existing methods in text fluency and factual consistency.

In summary, the main contributions of this paper are as follows:
\begin{itemize}

\item The proposal of an adaptive dual-watermarking framework combines the existence verification capability of zero-bit watermarking with the information tracing ability of multi-bit watermarking, achieving efficient and fine-grained content authentication within a unified framework.

\item The design of a dynamic hard-coded list mechanism builds upon the green list, dynamically partitions black and white lists using pseudo-random hash functions, and employs a hard-selection rule to unambiguously map message bits to vocabulary choices, enhancing the reliability and stealth of information embedding.

\item The introduction of an entropy-adaptive embedding strategy allows intelligent selection of embedding opportunities based on the predictive entropy of the language model, enforcing strong embedding in high entropy regions to ensure capacity while prioritizing text quality protection in low-entropy regions, thus achieving an optimal balance between watermark strength and text fluency.

\item The achievement of efficient long bit information embedding and extraction in short texts enables stable embedding of up to 64 bits of traceability information within paragraph-level short texts, addressing the limitation of insufficient embedding capacity in constrained paragraphs in existing technologies and providing a feasible authentication solution for practical scenarios such as food reports.
\end{itemize}

\section{Related Work}

\subsection{Zero-bit Watermarking}


Research on zero-bit watermarking has progressed from rule-based techniques to approaches leveraging modern language models. Early methods focused on rewriting-based strategies such as paraphrasing \cite{atallah2002natural} and synonym substitution \cite{topkara2006hiding}. With the rise of LLMs, more imperceptible watermarking became possible.
For example, Abdelnabi et al. \cite{abdelnabi2021adversarial} proposed using a reverse-trained text-to-text model to perform both insertion and extraction. He et al. \cite{he2022protecting, he2022cater} embedded watermarks via contextual lexical substitution.
A major milestone was the work by Kirchenbauer et al. \cite{kirchenbauer2023watermark}, who introduced a statistical watermark by modifying token-generation logits—an approach that has become a standard technique for distinguishing machine-generated from human-written text.

\subsection{Multi-bit Watermarking}



Multi-bit watermarking embeds information-bearing bit sequences into generated text, enabling applications such as source attribution and content tracking.
Fernandez et al. \cite{fernandez2023three} improved robustness using stronger statistical tests, while Yoo et al. \cite{yoo2024advancing} assigned short bit segments to tokens through pseudo-random mapping. Qu et al. \cite{qu2025provably} further enhanced robustness through pseudo-random segment assignment. However, a common limitation of the aforementioned schemes is their requirement for a large number of tokens to embed the watermark, indicating a lack of high-density embedding capability.

\section{Methodology}

\subsection{Overall Framework}

The ToSS framework proposed in this paper adopts a synergistic dual-layer watermarking architecture, aiming to achieve everything from efficient existence verification to fine-grained information tracing. Building upon the classic zero-bit watermarking foundation, this design innovatively introduces an encodable multi-bit watermarking layer and balances watermark strength with text quality through an adaptive mechanism. Its detailed design is illustrated in Fig. \ref{fig2}.
In particular, ToSS incorporates an entropy-adaptive decision module that distinguishes between high-entropy and low-entropy positions during text generation. As shown in the figure, high-entropy tokens (marked in red) represent regions where the model is uncertain and thus suitable for full-strength multi-bit watermark embedding through dynamic black–white list selection. Low-entropy tokens (marked in light blue), which typically correspond to factual or semantically constrained content, receive only a weak green-list bias and no multi-bit embedding, ensuring that text fluency and factual accuracy are preserved.

The existence verification layer of the watermark is based on the green list method using statistical hypothesis testing. This layer implants detectable statistical features into the text by biasing the generation probability of green list vocabulary during the text generation process. For a given text segment to be detected, one can efficiently determine whether it contains a watermark by calculating its Z-score:

\begin{equation}
z = \frac{|S_G| - \gamma T}{\sqrt{T \gamma (1 - \gamma)}}
\end{equation}
Comparing it with a preset threshold, where \( |S_G| \) is the number of green list tokens, \( T \) is the text length, and \( \gamma \) is the green list ratio.

The information tracing layer is the core innovation of this framework. It constructs an encodable communication channel based on the existence verification layer. This layer converts traceability information such as user identity and timestamp into a binary sequence \(\mathbf{m} \in \{0,1\}^L\) and embeds the information bits into the vocabulary selection pattern through a dynamic pseudo-random mapping mechanism. Specifically, at each generation timestep \(t\), based on the current message bit \(m_i\) to be embedded, the system further partitions the base green list \(G_t\) into a white list \(W_t\) and a black list \(B_t\), achieving information encoding by adjusting the sampling probability. Furthermore, an entropy-adaptive embedding strategy dynamically adjusts the watermark embedding strength \( \delta' = \alpha(H_t) \cdot \delta \) according to the predictive entropy value \( H_t \) of the language model at each generation position. This ensures the robustness of information embedding in high-uncertainty contexts while prioritizing text fluency and accuracy for low-uncertainty critical information.

This layered design allows ToSS to retain the detection efficiency of zero-bit watermarks while achieving the capability to embed long-bit traceability information in paragraph-level short texts. It provides a practical solution for scenarios requiring high information density authentication, such as food safety reports.

\begin{table*}[htbp]
\centering
\caption{Comparison of watermarking performance across representative methods under different bit lengths. }
\resizebox{\textwidth}{!}{%

\begin{tabular}{|c|c|c|c|c|c|c|c|c|c|c|c|c|c|c|c|}
\hline
\textbf{Method} & \multicolumn{3}{c|}{\textbf{12-bit}} & \multicolumn{3}{c|}{\textbf{16-bit}} & \multicolumn{3}{c|}{\textbf{20-bit}} & \multicolumn{3}{c|}{\textbf{24-bit}} & \multicolumn{3}{c|}{\textbf{32-bit}} \\
\cline{2-16}
& \begin{tabular}{@{}c@{}}Match\\(\%)\end{tabular} & \begin{tabular}{@{}c@{}}Bit Acc\\(\%)\end{tabular} & \begin{tabular}{@{}c@{}}Time\\(s)\end{tabular} & \begin{tabular}{@{}c@{}}Match\\(\%)\end{tabular} & \begin{tabular}{@{}c@{}}Bit Acc\\(\%)\end{tabular} & \begin{tabular}{@{}c@{}}Time\\(s)\end{tabular} & \begin{tabular}{@{}c@{}}Match\\(\%)\end{tabular} & \begin{tabular}{@{}c@{}}Bit Acc\\(\%)\end{tabular} & \begin{tabular}{@{}c@{}}Time\\(s)\end{tabular} & \begin{tabular}{@{}c@{}}Match\\(\%)\end{tabular} & \begin{tabular}{@{}c@{}}Bit Acc\\(\%)\end{tabular} & \begin{tabular}{@{}c@{}}Time\\(s)\end{tabular} & \begin{tabular}{@{}c@{}}Match\\(\%)\end{tabular} & \begin{tabular}{@{}c@{}}Bit Acc\\(\%)\end{tabular} & \begin{tabular}{@{}c@{}}Time\\(s)\end{tabular} \\
\hline
Fernandez et al. \cite{fernandez2023three} & 99.6 & 100.0 & 0.12 & 99.6 & 100.0 & 0.34 & 99.2 & 99.9 & 5.04 & 98.0 & 99.8 & 110 & NA & NA & \begin{tabular}{@{}c@{}}29000\\(Estimated)\end{tabular}   \\
\hline
Wang et al. \cite{wang2023towards} & 99.6 & 100.0 & 0.16 & 98.8 & 99.9 & 0.58 & 98.4 & 99.8 & 3.14 & 97.2 & 99.6 & 35.5 & NA & NA & \begin{tabular}{@{}c@{}}8300\\(Estimated)\end{tabular}  \\
\hline
Yoo et al. \cite{yoo2024advancing} & 86.4 & 96.5 & 0.01 & 73.6 & 94.6 & 0.01 & 49.2 & 90.8 & 0.01 & 30.4 & 89.4 & 0.01 & 8.4 & 81.2 & 0.01 \\
\hline
Cohen et al. \cite{cohen2025watermarking} & 93.2 & 98.7 & 0.01 & 88.8 & 97.0 & 0.02 & 78.4 & 95.3 & 0.02 & 65.6 & 94.7 & 0.03 & 27.2 & 89.7 & 0.04 \\
\hline
Qu et al. \cite{qu2025provably}& 98.8 & 99.9 & 0.04 & 98.0 & 99.7 & 0.06 & 97.6 & 99.6 & 0.10 & 96.0 & 99.5 & 0.18 & 94.0 & 99.1 & 0.6 \\
\hline
\textbf{ToSS(Ours)} & \textbf{100.0} & \textbf{100.0} & \textbf{0.35} & \textbf{99.45} & \textbf{99.93} & \textbf{0.36} & \textbf{99.43} & \textbf{99.89} & \textbf{0.36} & \textbf{98.95} & \textbf{99.79} & \textbf{0.37} & \textbf{98.95} & \textbf{99.65} & \textbf{0.39} \\
\hline
\end{tabular}
}
\label{tab:table1}
\end{table*}

\subsection{Key Ideas of Our Watermark Design}

To achieve reliable embedding of multi-bit information, this work proposes a dynamic list partitioning mechanism based on hard selection. Building upon the green list of traditional zero-bit watermarks, this mechanism constructs a deterministic information encoding channel through a pseudo-random hash function. Specifically, at each generation timestep \(t\), the system first generates a base green list \(G_t\) based on the previous token \(s_{t-1}\) and a key \(K\), with its size approximately \(\gamma \cdot |V|\), consistent with the standard procedure of such methods. Subsequently, based on the current message bit index \(msg\_index\) and the key \(K\), the system pseudo-randomly partitions the green list \(G_t\) into equally sized black list \(b\_list\) and white list \(w\_list\):

\begin{equation}
G_t = b\_list \cup w\_list,
\qquad
|b\_list| = |w\_list| = |G_t|/2
\end{equation}

Let the message to be embedded be 

\begin{equation}
\mathbf{m} = (m_1, m_2, \dots, m_L) \in \{0,1\}^L
\end{equation}

The hard-encoding rule ensures an unambiguous mapping from each message bit to vocabulary selection: when \(m_i = 0\), the next token is sampled uniformly from the black list \(b\_list\); when \(m_i = 1\), it is sampled uniformly from the white list \(w\_list\). 

Formally,


\begin{equation}
s_t =
\begin{cases}
\text{Uniform}(b\_list), & m_i = 0 \\
\text{Uniform}(w\_list), & m_i = 1
\end{cases}
\end{equation}

This deterministic binding eliminates the inherent uncertainty of traditional probability-bias watermarking, thereby forming a solid foundation for reliable extraction.

During the extraction phase, the detector reconstructs the list partition at each timestep using the same key \(K\). The embedded information bits are then recovered using maximum-likelihood decoding:

\begin{equation}
\hat{m}_i
=
\arg\max_{b \in \{0,1\}}
\sum_{t}
\mathbb{I}\big( s_t \in \mathcal{L}_t(b) \big)
\end{equation}

This dynamic mechanism inherits the detection efficiency of zero-bit watermarks while maintaining encoding stealth through pseudo-random list partitioning. However, we observe that the algorithm attempts to embed watermark bits at every generation position, which leads to substantial degradation in text quality. Therefore, we further develop an entropy-adaptive strategy that restricts such strong forced embedding to suitable high-entropy regions, thereby jointly optimizing both information reliability and text quality.



\subsection{Adaptive Entropy Embedding Algorithm}

To address the potential degradation of text quality caused by hard encoding in low-entropy regions, this work introduces an entropy-adaptive embedding strategy grounded in predictive uncertainty. The core idea is to leverage the entropy value  

\begin{equation}
H_t = -\sum_{v \in V} P_t(v) \log P_t(v)
\end{equation} 
of the language model at each generation step as a signal for dynamically adjusting the watermark embedding strength. When the model exhibits high prediction certainty (low-entropy regions), text quality is prioritized; when prediction uncertainty is high (high-entropy regions), strong watermark embedding is applied to ensure reliable information transmission.

Specifically, the system first sets an entropy threshold \(H_{th}\). At each generation timestep \(t\), the current entropy value \(H_t\) is computed and compared with the threshold. If \(H_t < H_{th}\), the watermark embedding operation is skipped, and the token \(s_t\) is sampled directly from the original probability distribution \(p_t\). In this case, the message-bit index \(msg\_index\) is not advanced. This mechanism ensures that the algorithm preserves text fluency in low-entropy regions, such as key terminology and fixed expressions.

When \(H_t \geq H_{th}\), the system enters the watermark embedding mode. If there are still message bits to be written (i.e., \(msg\_index < L\)), the system forcibly samples from the corresponding black or white list according to the current message bit \(m_i\). Unlike traditional hard-encoding watermarking, the bit index is advanced only after the bit has been successfully embedded, which significantly improves the completion rate of long-message embedding under limited text length.

After all message bits have been embedded (\(msg\_index \geq L\)), the system automatically falls back to a single-layer green-list biasing mode, where a bias \(\delta\) is applied to the logits of tokens in the green list \(G_t\) before sampling. This retains watermark existence-detection capability while avoiding unnecessary computational overhead.



\begin{figure}[t] 
\centering
  \includegraphics[width=1\linewidth]{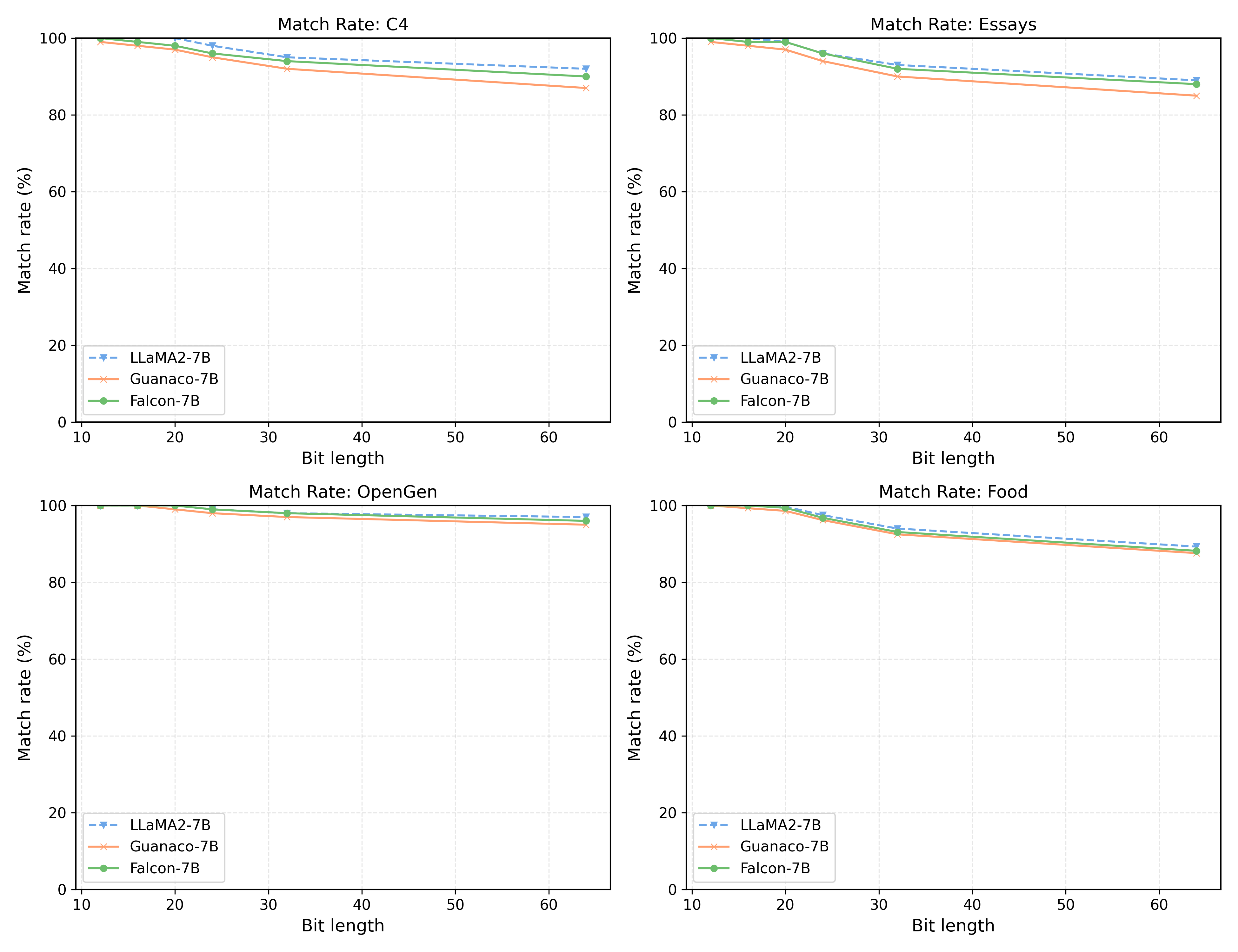} 
  \caption{Match rate performance of the proposed ToSS method across different large language models and datasets.
}   
  \label{fig3}
\end{figure}

\begin{figure}[t] 
\centering
  \includegraphics[width=1\linewidth]{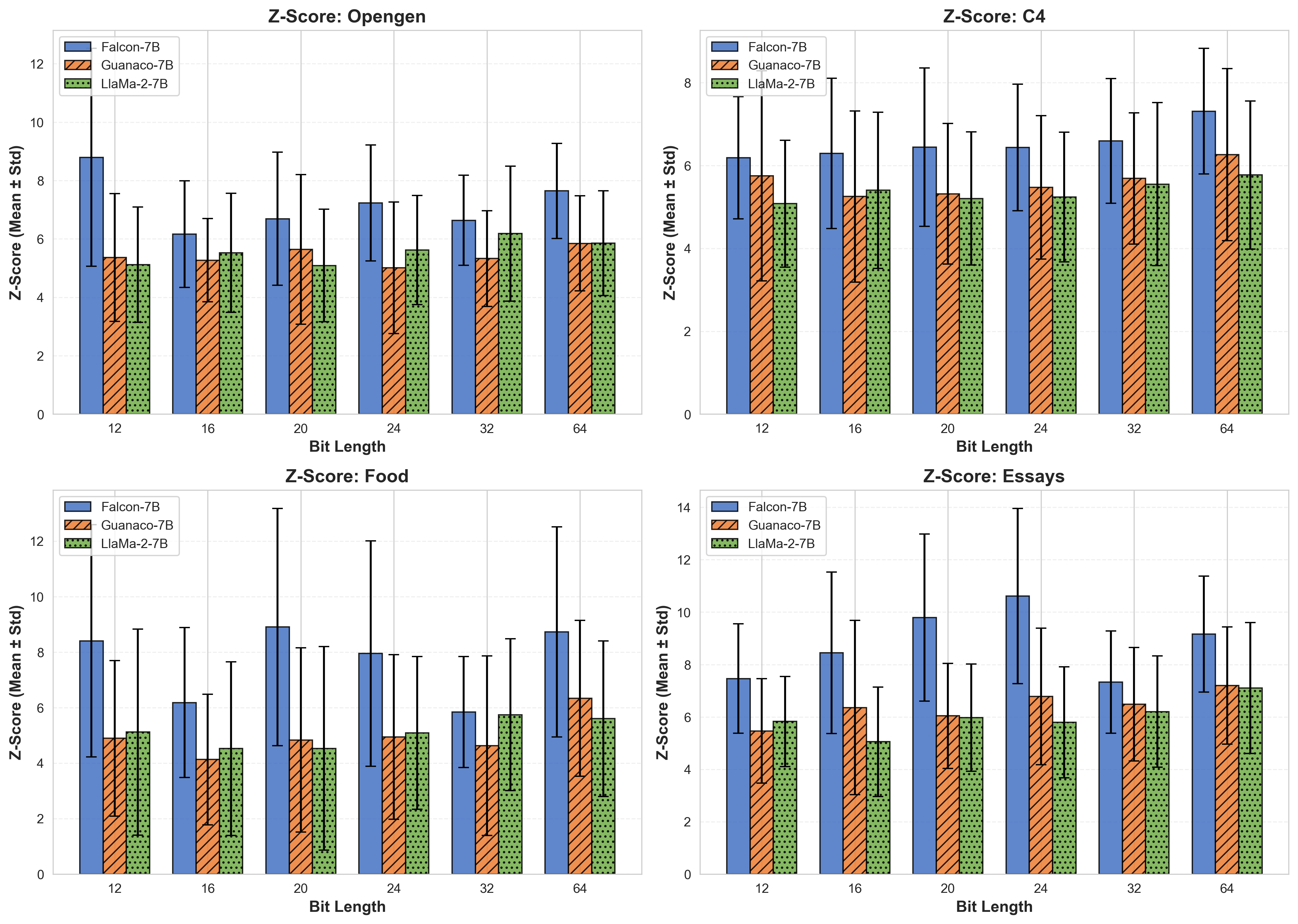} 
  \caption{Z-score performance of the ToSS watermark across different large language models and datasets.
}   
  \label{fig4}
\end{figure}

\section{Experiment}

\subsection{Experiment Setup}


\begin{table*}[h]
\centering
\caption{Performance of the LLaMA‑2‑7B model on the OpenGen dataset under different entropy thresholds}
\resizebox{\textwidth}{!}{%
\begin{tabular}{|c|c|c|c|c|c|c|c|c|c|c|c|c|c|c|c|c|c|c|}
\hline
\textbf{Threshold} & \multicolumn{3}{c|}{\textbf{12-bit}} & \multicolumn{3}{c|}{\textbf{16-bit}} & \multicolumn{3}{c|}{\textbf{20-bit}} & \multicolumn{3}{c|}{\textbf{24-bit}} & \multicolumn{3}{c|}{\textbf{32-bit}} & \multicolumn{3}{c|}{\textbf{64-bit}} \\
\cline{2-19}
& \begin{tabular}{@{}c@{}}Match\\(\%)\end{tabular} & \begin{tabular}{@{}c@{}}Bit Acc\\(\%)\end{tabular} & \begin{tabular}{@{}c@{}}Time\\(s)\end{tabular} & \begin{tabular}{@{}c@{}}Match\\(\%)\end{tabular} & \begin{tabular}{@{}c@{}}Bit Acc\\(\%)\end{tabular} & \begin{tabular}{@{}c@{}}Time\\(s)\end{tabular} & \begin{tabular}{@{}c@{}}Match\\(\%)\end{tabular} & \begin{tabular}{@{}c@{}}Bit Acc\\(\%)\end{tabular} & \begin{tabular}{@{}c@{}}Time\\(s)\end{tabular} & \begin{tabular}{@{}c@{}}Match\\(\%)\end{tabular} & \begin{tabular}{@{}c@{}}Bit Acc\\(\%)\end{tabular} & \begin{tabular}{@{}c@{}}Time\\(s)\end{tabular} & \begin{tabular}{@{}c@{}}Match\\(\%)\end{tabular} & \begin{tabular}{@{}c@{}}Bit Acc\\(\%)\end{tabular} & \begin{tabular}{@{}c@{}}Time\\(s)\end{tabular} & \begin{tabular}{@{}c@{}}Match\\(\%)\end{tabular} & \begin{tabular}{@{}c@{}}Bit Acc\\(\%)\end{tabular} & \begin{tabular}{@{}c@{}}Time\\(s)\end{tabular} \\
\hline
0.0 & 100.0 & 100.0 & 0.12 & 100.0 & 100.0 & 0.17 & 100.0 & 100.0 & 0.18 & 100.0 & 100.0 & 0.19 & 100.0 & 100.0 & 0.21 & 100.0 & 100.0 & 0.28 \\
\hline
1.0 & 100.0 & 100.0 & 0.35 & 99.45 & 99.93 & 0.36 & 99.43 & 99.89 & 0.36 & 98.95 & 99.79 & 0.37 & 98.95 & 99.65 & 0.39 & 95.63 & 99.12 & 0.47 \\
\hline
2.0 & 98.00 & 99.85 & 0.37 & 97.00 & 99.75 & 0.40 & 95.88 & 99.62 & 0.43 & 94.40 & 99.49 & 0.47 & 92.50 & 99.32 & 0.51 & 89.38 & 99.10 & 0.64 \\
\hline
3.0 & 98.0 & 98.0 & 0.36 & 88.0 & 93.3 & 0.37 & 90.0 & 90.0 & 0.42 & 84.0 & 86.4 & 0.42 & 66.0 & 72.4 & 0.46 & 16.0 & 17.4 & 0.47 \\
\hline
4.0 & 74.0 & 75.2 & 0.44 & 58.0 & 59.6 & 0.45 & 44.0 & 46.4 & 0.46 & 40.0 & 40.0 & 0.47 & 16.0 & 16.0 & 0.47 & 0.0 & 0.0 & 0.49 \\
\hline
\end{tabular}%
}
\label{tab:table2}
\end{table*}

\begin{table}[h]
\centering
\caption{Effect of different entropy thresholds on text perplexity (PPL) of the LLaMA‑2‑7B model on the OpenGen dataset}
\begin{tabular}{|c|c|c|c|c|c|c|}
\hline
\textbf{Threshold} & \textbf{12-bit} & \textbf{16-bit} & \textbf{20-bit} & \textbf{24-bit} & \textbf{32-bit} & \textbf{64-bit} \\
\hline
0.0 & 4.82 & 5.26 & 4.88 & 5.11 & 6.41 & 9.99 \\
\hline
1.0 & 4.32 & 4.29 & 4.52 & 4.53 & 5.20 & 7.14 \\
\hline
2.0 & 4.48 & 4.54 & 4.62 & 4.73 & 4.89 & 5.31 \\
\hline
3.0 & 4.00 & 3.90 & 3.74 & 4.02 & 3.81 & 4.28 \\
\hline
4.0 & 3.80 & 3.87 & 3.90 & 4.00 & 3.77 & 3.84 \\
\hline
\end{tabular}%
\label{tab:table3}
\end{table}

Following  prior research\cite{qu2025provably}, we evaluate our method ToSS using the following datasets:  OpenGen\cite{krishna2023paraphrasing}, C4-News\cite{raffel2020exploring}, and Essays\cite{schuhmann2023essays}. Specifically, OpenGen contains 3,000 two-sentence passages randomly sampled from the WikiText103\cite{merity2016pointer} validation set. The C4-News dataset comprises roughly 15GB of online news text. The Essays dataset consists of student assignments from the IvyPanda repository\cite{ivypanda2024essays}. To better assess watermarking in food safety applications, we also introduce a new food-related dataset\cite{corbt_all_recipes}. It includes diverse information such as recipes, ingredient lists, and cooking instructions. Unless specified otherwise, we primarily use the widely adopted OpenGen dataset as our main benchmark, consistent with prior work\cite{qu2025provably}.

Our experiments primarily utilize the following state-of-the-art publicly available LLMs: LLaMA-2-7B\cite{ zhao2023provable}, Guanaco-7B\cite{dettmers2023qlora }, and Falcon-7B\cite{almazrouei2023falcon}. By default, and consistent with prior work \cite{qu2025provably}, we employ LLaMA-2-7B as our baseline model.

We employ the following metrics for evaluation: match rate to assess the capability of fully extracting watermark information, and bit accuracy to reflect the precision of information retrieval. For quality assessment, we report both the raw perplexity and the Z-score derived after watermark decoding to measure text quality and watermarking effectiveness, respectively.

Additionally, we test our method with watermark bit lengths of 12, 16, 20, 24, 32, and 64 bits, setting the entropy threshold to 1.0 by default while maintaining the default values for all other parameters.

\subsection{Comparison with Other Methods}

We conducted a thorough comparison between our method and existing representative watermarking schemes, with particular focus on the recent approach by Qu et al. \cite{qu2025provably}. Experiments were performed under identical settings, measuring match rate, bit accuracy, and extraction time across bit lengths from 12 to 32 bits. The results are summarized in Table \ref{tab:table1}.

In terms of decoding accuracy, our method achieved leading performance across all tested bit lengths. Even with 32-bit long-sequence embedding, it maintained a stable match rate of 98.95\% and bit accuracy of 99.65\%, showing minimal performance degradation. In comparison, while Qu et al. \cite{qu2025provably} achieved an excellent 94.0\% match rate at 32 bits, our method still held a nearly 5 percentage point advantage. The methods by Yoo et al. \cite{yoo2024advancing} and Cohen et al. \cite{cohen2025watermarking} suffered significant performance drops with long-bit embedding, with their 32-bit match rates plummeting to 8.4\% and 27.2\% respectively. This comparison strongly validates the effectiveness of our proposed dynamic list partitioning and entropy-adaptive mechanism in maintaining reliability for long-sequence information embedding. This stable retention capability is crucial for scenarios like food safety reports that require embedding complex traceability information such as production batch numbers, inspection timestamps, and organization codes.

Regarding extraction efficiency, our method demonstrated excellent overall performance. Its extraction time was on the same order of magnitude as Qu et al. \cite{qu2025provably} and remained stable across all bit lengths, confirming the controllable computational overhead of our approach. This sub-second detection speed meets the practical need for fast, on-the-fly verification of report authenticity within food supply chains. More importantly, our method showed a substantial efficiency advantage over the approaches of Fernandez et al. \cite{fernandez2023three} and Wang et al. \cite{wang2023towards}. For 24-bit embedding, our extraction time was 0.37 seconds, whereas the methods of \cite{fernandez2023three} and \cite{wang2023towards} surged to 110 seconds and 35.5 seconds respectively due to computational complexity, rendering them impractical for 32-bit embedding.

Notably, our method showed no significant performance degradation with long-sequence watermark embedding. As shown in Fig. \ref{fig3}, its watermark extraction accuracy remained high even when extended to 64-bit tests, significantly outperforming comparative methods. Specifically, on the Food dataset constructed to reflect the textual characteristics of the food domain, our method also achieved excellent and stable performance (shown in Fig. \ref{fig3} and Fig. \ref{fig4}), verifying its strong adaptability to domain-specific terminology and writing styles. Furthermore, the watermark evidence strength (Z-score) remained consistently high and stable across different watermark lengths, as shown in Fig. \ref{fig4}. This clearly indicates that our entropy-adaptive strategy effectively balances embedding strength and length, providing technical assurance for reliable food information traceability and authentication.

In summary, our method offers further improvement in decoding accuracy over the recent Qu et al. \cite{qu2025provably} approach, while far surpassing the computational efficiency of earlier high-accuracy schemes like \cite{fernandez2023three} and \cite{wang2023towards}. It achieves a superior overall balance between decoding accuracy, extraction efficiency, and long-sequence embedding capability, laying a solid foundation for building a high-assurance AI content authentication system for food safety.

\subsection{Ablation Study}





To test how well our entropy-based method works in critical areas like food safety, we carefully checked how different settings affect watermark performance and text quality. As shown in Tables \ref{tab:table2} and \ref{tab:table3},  when we set the threshold to 0.0, meaning the system always embeds watermarks regardless of content, watermark detection works perfectly. However, this comes at a clear cost to text quality. With longer watermarks, like 64 bits, the text becomes much less natural, with a perplexity score rising to 9.99. This shows that forcing watermarks into all parts of the text, including critical factual statements, can harm readability, which is unacceptable for professional reports.

For high-risk uses like food safety reports or nutrition guides, where traceability must be perfectly accurate, we set the default entropy threshold to 1.0. This ensures extremely reliable watermark extraction. Even with 64-bit watermarks, the system achieves a 95.63\% match rate and 99.12\% bit accuracy, allowing error-free recovery of key details like sample IDs or ingredient amounts. Text quality remains acceptable, with a perplexity of 7.14, balancing reliability and readability.

In situations where text fluency is more important, a threshold of 2.0 can be used. This improves text quality further (perplexity 5.31), while watermark performance (89.38\% match rate at 64 bits, 99.10\% bit accuracy) stays sufficient for most verification needs. This shows the flexibility of our approach. However, for high-risk fields, using thresholds above 2.0 is not recommended, as decoding performance drops sharply, undermining the watermark's core purpose. Therefore, a default threshold of 1.0 offers the best balance, providing strong, reliable content protection for applications involving public health and safety.

\section{Conclusion}


%

In summary, our proposed ToSS  framework offers a new technical approach for securing AI-generated content, particularly in high-risk fields like food safety. The core design of ToSS is highly sensitive to tampering. Any unauthorized modification to the text disrupts the densely embedded watermark information, providing a strong mechanism for verifying content integrity. This ``zero-tolerance" mechanism ensures text originality, which is a critical feature for applications requiring absolute trustworthiness. Experimental results confirm that ToSS achieves this goal while maintaining high decoding accuracy and practical usability. Our work provides a new direction for content authentication, and future research may explore adaptive frameworks that balance integrity sensitivity and editing robustness across different application scenarios.

\section{Acknowledgment}
Zhongli Fang, Yiran Chen, Lingyun Zhang, Yu Liu and Ping Chen were supported by National Key R\&D Program of China under grant No. 2023YFB3107404 and  the Key Research and Development Programme of Ningbo’s “Science and Technology Innovation Yongjiang 2035” Plan under grant No. 2025Z054.

\bibliographystyle{IEEEbib}
\bibliography{icme2026references}

\end{document}